\documentclass{article}

\usepackage{microtype}
\usepackage{graphicx}
\usepackage{subfigure}
\usepackage{booktabs} 

\usepackage{hyperref}

\newcommand{\Lya}{Ly-$\alpha$ }

\usepackage[accepted]{icml2025}

\usepackage{amsmath}
\usepackage{amssymb}
\usepackage{mathtools}
\usepackage{amsthm}

\usepackage[capitalize,noabbrev]{cleveref}

\theoremstyle{plain}

\theoremstyle{definition}

\theoremstyle{remark}

\usepackage[textsize=tiny]{todonotes}

\icmltitlerunning{BaryonBridge}

\begin{document}

\twocolumn[
\icmltitle{\texttt{BaryonBridge}: Stochastic Interpolant Model for Fast Hydrodynamical Simulations}

\icmlsetsymbol{equal}{*}

\begin{icmlauthorlist}
\icmlauthor{Benjamin Horowitz}{ipmu,cd3}
\icmlauthor{Carolina Cuesta-Lazaro}{cfa,iaifi,mit}
\icmlauthor{Omar Yehia}{ipmu,cd3}

\end{icmlauthorlist}
\icmlaffiliation{ipmu}{Kavli IPMU (WPI), UTIAS, The University of Tokyo, Kashiwa, Chiba 277-8583, Japan
}
\icmlaffiliation{cfa}{Harvard-Smithsonian Center for Astrophysics, 60 Garden Street, Cambridge, MA 02138, USA}
\icmlaffiliation{iaifi}{The NSF AI Institute for Artificial Intelligence and Fundamental Interactions, Massachusetts Institute of Technology, Cambridge, MA 02139, USA}
\icmlaffiliation{mit}{Department of Physics, Massachusetts Institute of Technology, Cambridge, MA 02139, USA}
\icmlaffiliation{cd3}{Center for Data-Driven Discovery, Kavli IPMU (WPI), UTIAS, The University of Tokyo, Kashiwa, Chiba 277-8583, Japan}

\icmlcorrespondingauthor{Benjamin Horowitz}{horowitz.ben@gmail.com}

\icmlkeywords{Machine Learning, ICML}

\vskip 0.3in
]



\printAffiliationsAndNotice{} 

\begin{abstract}
Constructing a general-purpose framework for mapping between dark matter simulations and observable hydrodynamical simulation outputs is a long-standing problem in modern astrophysics. In this work, we present a new approach utilizing stochastic interpolants to map between cheap fast particle mesh simulations and baryonic quantities in three dimensions, requiring a total of $7$ GPU minutes per 256$^3$ grid size simulation. Using the CAMELS multifield dataset, we are able to condition our mapping on both cosmological and astrophysical properties. We focus this work on hydrodynamical quantities suitable for \Lya observables finding excellent agreement up to small spatial scales, $k \sim 10.0$ ($h^{-1}$ Mpc) at $z=2.0$, for \Lya flux statistics. Our approach is fully convolutional, allowing training on comparatively small volumes and application to larger volumes, which was tested on TNG50.
\end{abstract}

\section{Introduction}
\label{Introduction}
Understanding the large-scale structure of the universe requires modeling both the nonlinear evolution of the dark matter cosmic web and the complex baryonic physics behind observable tracers. Cosmological hydrodynamical simulations have become essential tools for capturing these effects, but their high computational cost limits their use for next-generation surveys that demand large volumes and high fidelity. Estimating statistical quantities like power spectra and their covariances across wide parameter ranges becomes especially expensive, motivating the development of efficient surrogate models.

A parallel line of development has focused on the reconstruction of initial conditions (ICs) from late-time observations via forward modeling. These approaches, often relying on differentiable structure formation models, enable constrained simulations that match observed data and support field-level inference. While powerful, their effectiveness is currently limited by the lack of efficient differentiable models that incorporate baryonic physics. Bridging this gap is critical for enabling more realistic reconstructions that account for feedback processes and their impact on observables.

Recently, deep learning-based surrogate models have emerged to accelerate simulation workflows. These methods typically either (1) augment low-cost dark matter simulations with baryonic effects \citep{2022ApJ...929..160H,2022hyphy}, or (2) generate full hydrodynamical outputs directly \citep{2019MNRAS.487L..24T,2025MNRAS.538.1201B}. 

Our work builds on these by introducing; (1) A differentiable and efficient simulation backbone using \texttt{fastPM} \citep{2016MNRAS.463.2273F}, enabling integration into inverse modeling frameworks \citep{2017JCAP...12..009S,2022arXiv221109958L,2018JCAP...10..028M,2019TARDISI,2021TARDISII}, (2) a fully convolutional architecture that generalizes to arbitrary volume sizes while training on smaller boxes \citep{2022ApJ...929..160H,2022hyphy} (3) conditional modeling on cosmological and astrophysical parameters, crucial for capturing the effects of baryonic feedback and reducing uncertainties in cosmological inferences.

Unlike unconditional generative models \citep{2019arXiv190412846Z,2023arXiv231100833H,2024arXiv240210997A}, our conditional framework leverages dark matter inputs to ensure generalization and independence across realizations---features necessary for covariance estimation and field-level reconstruction tasks.
\section{Simulations}
\label{sec:sim}

\begin{figure*}
    \centering
    \includegraphics[trim={0 8.8cm 0 3.3cm},clip,width=0.95\linewidth]{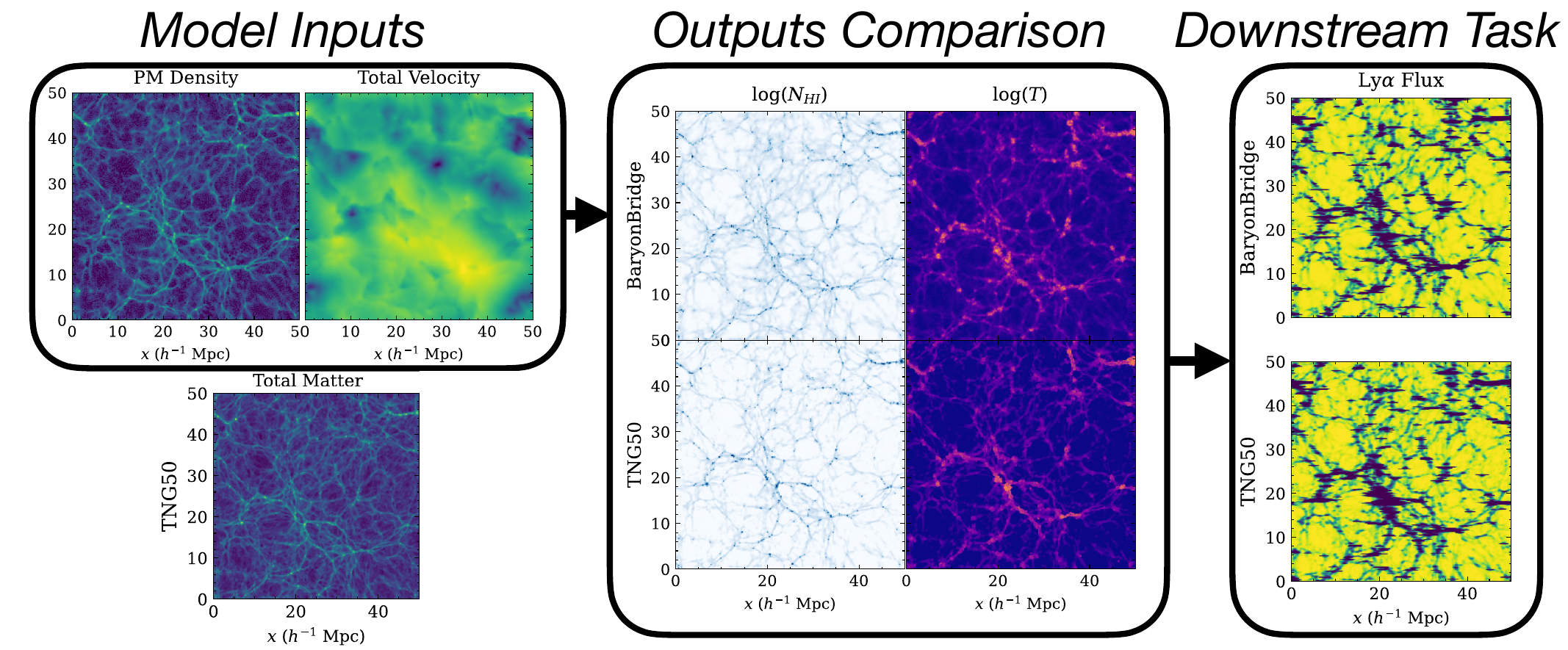}
    \vspace{-10pt}
    \caption{We show the input density \texttt{FastPM} generated density field and generated neutral hydrogen ($N_{HI}$), temperature (T), and inferred \Lya flux from our BaryonBridge model compared to the simulated true result in the \texttt{TNG50} simulation.}
    \label{fig:TNGPanel}
\end{figure*}

\subsection{CAMELS and TNG50}
CAMELS \citep{2021ApJ...915...71V} is a large-scale simulation effort designed to connect cosmological structure formation with astrophysical feedback processes using both N-body and hydrodynamic simulations. In this work, we focus exclusively on the IllustrisTNG suite of CAMELS, which employs the moving-mesh \texttt{AREPO} code \citep{2010MNRAS.401..791S,2020ApJS..248...32W} and adopts the same subgrid physics model as the original IllustrisTNG simulations \citep{2017MNRAS.465.3291W,2018MNRAS.473.4077P}. 

Each simulation in the IllustrisTNG suite follows the coupled evolution of dark matter and baryonic matter in a $(25~h^{-1}\mathrm{Mpc})^3$ periodic volume, using $256^3$ particles per component. Within this suite, the \textit{Latin Hypercube (LH)} set consists of 1,000 distinct simulations sampling a wide range of cosmological ($\Omega_m$ and $\sigma_8$) and astrophysical feedback ($A_{SN,1}$, $A_{SN,2}$, $A_{AGN,1}$, $A_{AGN,2}$) parameters using a Latin hypercube design. This enables broad coverage of the parameter space with relatively few samples.

We use the CAMELS Multifield Dataset (CMD) \citep{2022ApJS..259...61V}, which provides 3D grids derived from the CAMELS simulation outputs. Specifically, we specialize to the 3D volumes from the IllustrisTNG-LH set, leveraging its diversity in cosmological and feedback parameters to train and validate our model. For this work we focus on \Lya Forest observables, so we are interested in neutral hydrogen density, temperature, and line of sight (LOS) velocity. The grid LOS velocity is not available in the CMD dataset, so we construct it from available particle data.

To test our model on larger simulation boxes, we also use the \texttt{TNG50} simulation \citep{2019MNRAS.490.3234N}. Like the CAMELS boxes, it uses \texttt{AREPO} hydrodynamics code along with a TNG subgrid physics model. However, it has $\sim 3$ times the volume (35 $h^{-1}$ Mpc) at significantly higher particle density of $2160^3$. This change of both simulation volume and resolution will provide a robust test for the generalization of our model. 

\subsection{Particle Mesh Simulations}

For our conditioning field, we use particle mesh simulations with matched initial conditions of the full hydrodynamical simulations. We use a \texttt{JAX} implementation \cite{2022mla..confE..60L} of the \texttt{FastPM} algorithm \cite{2016MNRAS.463.2273F}. Fast particle mesh algorithms have become a standard tool for field level inference since they are straightforward to take analytical derivatives of output quantities with respect to initial conditions, this enables derivative based optimization/sampling methods for initial condition inference.\cite{2017JCAP...12..009S,2018JCAP...10..028M,2019TARDISI,2021TARDISII}

For our particle mesh simulations, we take the particle/velocity initial conditions given in the CAMELS simulation set and use them as ICs for our particle mesh code. We down-sample the particles uniformly to $256^3$, and use a force resolution $B=2$ to provide more accurate diffuse large scale structure necessary for \Lya statistics. We use a CIC painting scheme to readout the particle masses, total velocities, and line of sight velocities on a $128^3$ grid to match those in the CAMELS multifield set. We use the matter density and total velocity as inputs to our interpolant model.

\subsection{Mapping to \Lya Forest}
\label{subsec:lya}
For our example observable, we focus on \Lya forest since it is a field-level quantity which has high dependance on both astrophysics \cite{2022constrainfgpa} and cosmology \cite{Lukic2014}. The expression for optical depth, $\tau$, in real space, as given in \cite{Lukic2014}, is:
\begin{equation}
\tau_\nu
\equiv \int n_{HI} \sigma_\nu dr .\label{eq:tau_definition}
\end{equation}
\begin{figure}
    \centering
    \includegraphics[width=0.95\linewidth]{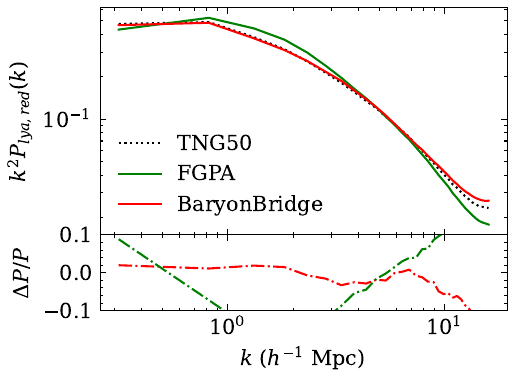}
    \vspace{-18pt}
    \caption{Comparison of redshift space \Lya flux for both the original \texttt{TNG50} simulation and that mapped via \texttt{BaryonBridge} code. We also compare to an analytical approximation given via the Flucuating Gunn-Peterson Approximation (FGPA) on the original dark matter field.}
    \label{fig:TNGPS}
\end{figure}
In this expression, $\nu$ denotes the frequency, $\sigma_\nu$ is the cross-section for the \Lya transition, and $dr$ represents an element of proper distance. When adopting a Doppler-broadened line profile—characterized by a Gaussian shape—the expression becomes:
\begin{equation}
\tau_\nu =
\frac{\pi e^2}{m_e c} f_{12} \int \frac{n_{HI}}{\Delta \nu_D} \frac{\exp \left[-\left(\frac{\nu - \nu_0}{\Delta \nu_D}\right)^2 \right]}{\sqrt{\pi}} dr,
\end{equation}\label{eq:tau_real}
Here, $e$ is the elementary charge, $m_e$ is the electron mass, and $c$ is the speed of light. The quantity $\nu_0$ refers to the central frequency of the transition, $f_{12}$ is the oscillator strength for the  \Lya resonance, and $\Delta \nu_D = (b/c) \nu_0$ is the Doppler width, with the Doppler parameter defined as $b = \sqrt{2k_B T/m_H}$. We use a JAX version of the differentiable GPU-accelerated code \texttt{THALAS} \citep{2024thalas} to calculate these quantities from simulations.

\section{Stochastic Interpolant Models}
\label{sec:stochastic}

We employ Stochastic Interpolants, as formulated in \citep{albergo2023stochasticinterpolantsunifyingframework, chen2024probabilisticforecastingstochasticinterpolants} and first applied in the cosmological context in \citep{cuestajoint}, to model the conditional distribution $p(\delta_\mathrm{baryons}|\delta_\mathrm{dark matter}, \mathcal{C}, \mathcal{A})$. Given a dark matter density field, $\delta_{\mathrm{dark matter}}$, and a set of cosmological and astrophysical parameters $\mathcal{C}$, $\mathcal{A}$, (in our case, 
$\Omega_m,\sigma_8,A_{SN1},A_{SN2},A_{AGN1},A_{AGN2}$) we seek to generate a probabilistic ensemble of possible baryon distributions $\delta_\mathrm{baryons}$. A stochastic interpolant $I_s$ can be constructed as:
\begin{equation}
\label{eq:interpolant}
    X_s = \alpha_s x_0 + \beta_s x_1 + \sigma_s W_s,
\end{equation}
where $W_s$ is a Wiener process that can be sampled as $W_s = \sqrt{s}z$ with $z \sim \mathcal{N}(0, I)$. The interpolant maps a point mass measure at $x_0$ to the conditional distribution $p(x_s|x_0)$ as $s$ varies from $0$ to $1$. In our application, $x_0 = \delta_\mathrm{dark matter}$ and $x_1 = \delta_\mathrm{baryons}$.

The interpolant satisfies boundary conditions $\alpha_0 = \beta_1 = 1$ and $\alpha_1=\beta_0=\sigma_0 = 0$. For our implementation, we choose $\alpha_s = 1 - s$, $\beta_s = s^2$, and $\sigma_s = 1 - s$.

The conditional distribution defined by the interpolant, $p(I_s|x_0)$, corresponds to the law of a solution to a stochastic differential equation (SDE) that can be used as a generative model \citep{chen2024probabilisticforecastingstochasticinterpolants}:
\begin{equation}
    dX_s = b_s(X_s, x_0) ds + \sigma_s dW_s, \quad X_{s=0} = x_0,
\end{equation}
where the drift term $b_s$ is learned by parameterizing a neural network $\hat{b}_s(x,x_0,s)$ and minimizing the objective:
\begin{equation}
    L_b \left[\hat{b}_s \right] = \int_0^1 ds \mathbb{E} \left[ | \hat{b}_s(I_s, x_0, s) - R_s |^2 \right].
\end{equation}

Here, $R_s$ is determined by the interpolant as $R_s = \dot{\alpha}_s x_0 +\dot{\beta}_s x_1 + \dot{\sigma}_s W_s$, where the dots represent derivatives with respect to $s$.

Unlike conditional diffusion models, like the ones used for mapping dark matter fields to galaxies \citep{bourdin2024inpaintinggalaxycountsnbody}, stochastic interpolants leverage a base distribution $x_0$ that is already structurally similar to the target distribution, simplifying the mapping that the neural network needs to learn.

We implement the drift term using a fully convolutional 3D U-Net architecture. The fully convolutional nature allows us to train the model on small-volume hydrodynamical simulations (e.g., CAMELS) while performing inference on larger volumes. The interpolant time parameter $s$ is encoded together with the cosmological and astrophysical parameters and incorporated into the network input alongside $X_s$.

\section{Results}

We have trained our network, described in Sec. \ref{sec:stochastic} to map from the fast particle mesh simulations to the CAMELS simulations described in Sec. \ref{sec:sim}. We then map our output baryon fields to the \Lya flux statistics described in Sec. \ref{subsec:lya}. For redshift space distortions, we use the dark matter line of sight velocities from the particle mesh without machine learning augmentation.

\subsection{Inference over larger volumes: TNG50}

For a robust test of the utility of our trained model, we run a particle mesh simulation with matched initial conditions to that of the \texttt{TNG50} simulation and calculate a binned density and total velocity field. We apply our trained \texttt{BaryonBridge} on this volume. From the output $N_{HI}$ and $T$ fields, combined with the line of sight velocity from the particle mesh dark matter simulation, we calculate the \Lya forest as described in Sec \ref{subsec:lya}. We show our qualitative performance in a slice in Figure \ref{fig:TNGPanel}, finding excellent visual agreement, including predictions of thermal shocks which were difficult to to recover in past machine-learning based approaches \citep{2022ApJ...929..160H}, although with some loss of sharpness on the edges. We show the calculated \Lya power spectra in Figure \ref{fig:TNGPS}, finding excellent quantitative agreement with less that 3 \% error up to a $k \sim 8$ $h$ Mpc$^{-1}$. 

In total, our model requires only $7$ GPU-minutes for inference, one minute on four GPUs for particle mesh and three minutes on one GPU for a model sample, and 96 GPU-hours for training, compared to TNG50's $130$ million CPU core hours.

For an analytical comparison, we use the Fluctuating-Gunn Peterson Approximation for the temperature and baryon density field as described in \citet{2022ApJ...929..160H}. This method goes beyond the ``standard" FGPA method used (e.g. that used in \citet{2019TARDISI,2022constrainfgpa}) by predicting a temperature and density field with an exponential mapping, and then passing through an equation of state code to calculate the neutral hydrogen fraction and line of sight integration code.
\subsection{Parameter variations: CAMELS Boxes}

We also test our trained \texttt{BaryonBridge} model on validation boxes in the original IllustrisTNG-LH set held back from training. This set spans a range of cosmological and astrophysical properties, allowing us to validate our conditional mapping. Across our conditioning set, we find high accuracy in our output quantities of $N_{HI}$ and $T$ in terms of quantity power spectra and cross correlation coefficient. In particular, we find $<10\%$ error in the power spectra across the validation set up to a $k \sim 10$ $h$ Mpc$^{-1}$ as well as a correlation coefficient $>0.9$ up to a $k \sim 4$ $h$ Mpc$^{-1}$ for both of these fields.

\begin{figure}
    \centering

    \includegraphics[width=0.90\linewidth]{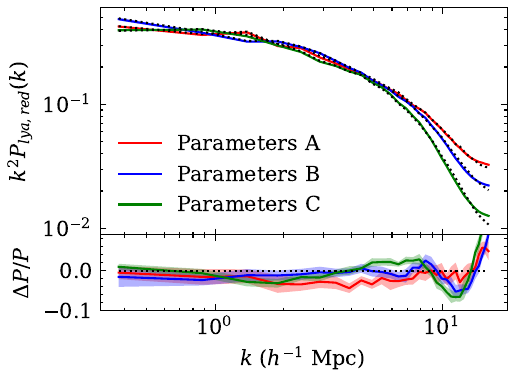}
    \vspace{-16pt}
    
    \caption{Power spectra statistics corresponding to the models in Tab. \ref{param_table}. The shaded region shows the variance of independent samples drawn from the same initial conditions; i.e. a model prediction of uncertainty in the underlying mapping.}
    \label{fig:camels_ps}
\end{figure}

We select a few extreme values, shown in Table \ref{param_table}, to further analyze in terms of the downstream task of \Lya analysis. We show the \Lya power spectra in Fig \ref{fig:camels_ps}, finding excellent agreement, with less than 5\% error up to a $k=10$ $h$ Mpc$^{-1}$ across the parameter space. The worst performing validation case, Parameters C, is quite extreme in both cosmological and astrophysical parameters, but still has acceptable error properties. We have also calculated the cross correlation coefficent, finding $r>0.9$ up to $k \sim 5.0$ $h$ Mpc$^{-1}$ across all validation parameter sets.

\begin{table}[]
\label{param_table}
\centering
\small
\begin{tabular}{l||llllll||l}
 & $\Omega_m$ & $\sigma_8$ & $A_{S1}$ & $A_{S2}$ & $A_{A1}$ & $A_{A2}$ & $\frac{\Delta\langle F \rangle}{\langle F \rangle}$ \\
 \hline
A & 0.46 & 0.87 & 3.81 & 0.70 & 1.50 & 1.78 & 0.2\% \\
B & 0.22 & 0.79 & 0.82 & 0.27 & 0.55 & 1.11 & 0.2\% \\
C & 0.14 & 0.68 & 2.34 & 2.83 & 0.79 & 0.54 & 0.3\%
\end{tabular}
\caption{Validation parameter sets used to demonstrate our conditional model. We also show the relative error on the inferred mean \Lya flux.}
\end{table}

\section{Discussion and Conclusion}

In this work, we have demonstrated a fast and accurate way to map from relatively low-resolution dark matter simulations to full hydrodynamical quantities in the case of \Lya observables. We have trained our methods on relatively small (25 $h^{-1}$ Mpc)$^3$ boxes from the CAMELS set and tested them on the larger \texttt{TNG50} box, finding excellent performance on downstream tasks like \Lya power spectra analysis. This method can be used not only for fast generation of mock catalogs for next generation surveys, but could also play a critical role in dynamical forward modeling field level approaches.

Remaining variance between \texttt{BaryonBridge} and the CAMELS results could in part be caused by poor resolution in low-mass regions from the underlying \texttt{JaxPM} simulation and full dark matter simulations. These difference could in part be alleviated via (conditional) refinement methods or neural augmentations on the particle mesh methods \cite{2022mla..confE..60L,2023arXiv231118017P}. 

In this work, we have focused on neutral hydrogen and temperature as they are required quantities for \Lya flux statistics. However, in preliminary tests we have found no fundamental restriction on target fields available in the CAMELS multifield dataset. In addition, thematically similar techniques have been used to map to highly non-linear quantities like stellar mass \cite{2024ApJ...970..174O}. Currently, memory limitations and training-time expediency require some specialization to only a few target fields at once. 

We calculate the \Lya statistics from the predicted gas temperature and density fields rather than directly inferring the transmitted flux for several key reasons. First, predicting the underlying physical fields enables us to study correlations with other observables. Second, this approach transforms the problem into a local mapping between the dark matter distribution and gas properties, avoiding the need to model the complex long-range correlations that arise when targeting the \Lya field directly in redshift space. Finally, the causal relationship between baryonic feedback processes and the observed \Lya absorption is more naturally captured by modeling their separate effects on the temperature and density fields, which then combine to determine the Lyman alpha field.

It is notable that all steps in our analysis framework, \texttt{JaxPM} particle mesh outputs to \texttt{BaryonBridge} baryon quantities to \texttt{THALAS} \Lya flux are differentiable allowing an end-to-end pipeline for explicit field level inference (e.g. \citet{2019TARDISI,2019A&A...630A.151P}). We will further explore this application in future works.

\bibliography{example_paper, mypapers,example}
\bibliographystyle{icml2025}

\onecolumn



\end{document}